\documentclass[aps,prl,twocolumn]{revtex4-1}

\usepackage{lipsum}
\usepackage{graphicx}
\usepackage{subfigure}
\usepackage{color}
\usepackage{amsmath}
\usepackage[linktocpage,colorlinks=true,linkcolor=blue,citecolor=blue,urlcolor=blue,breaklinks=true]{hyperref}
\usepackage{verbatim}
\usepackage{breakcites}
\usepackage{wrapfig}
\usepackage{soul}
\usepackage{bbold}
\usepackage{array,multirow}

\begin{document}
\title{Photo-bioconvection: towards light-control of flows in active suspensions} 

\author{A. Javadi$^{1}$, J. Arrieta$^{2}$, I. Tuval$^{2,3}$, and M. Polin$^{1,4}$}

\affiliation{$^{1}$ Physics Department, and $^4$Centre For Mechanochemical Cell Biology, University of Warwick, Gibbet Hill Road, Coventry CV4 7AL, United Kingdom  \\
	$^{2}$Instituto Mediterr\'aneo de Estudios Avanzados, IMEDEA, UIB-CSIC, 07190, Esporles, Spain\\
	$^{3}$Departamento de F\'isica, Universitat de les Illes Balears, 07122 Palma de Mallorca, Spain}

%

\email[Correspondence: ]{M.Polin@warwick.ac.uk}

\begin{abstract}
The persistent motility of the individual constituents in microbial suspensions represents a prime example of so-called active matter systems. Cells consume energy, exert forces and move, overall releasing the constraints of equilibrium statistical mechanics of passive elements and allowing for complex spatio-temporal patterns to emerge. Moreover, when subject to physico-chemical stimuli their collective behaviour often drives large scale instabilities of hydrodynamic nature, with implications for biomixing in natural environments and incipient industrial applications. In turn, our ability for external control of these driving stimuli could be used to govern the emerging patterns. Light, being easily manipulable and, at the same time, an important stimulus for a wide variety of microorganisms, is particularly well suited to this end. In this paper, we will discuss the current state, developments, and some of the emerging advances in the fundamentals and applications of light-induced bioconvection with a focus on recent experimental realisations and modelling efforts.
\end{abstract}


\maketitle
\section{Introduction}
\begin{figure}[b]
\centering
		\includegraphics[width=0.95\linewidth]{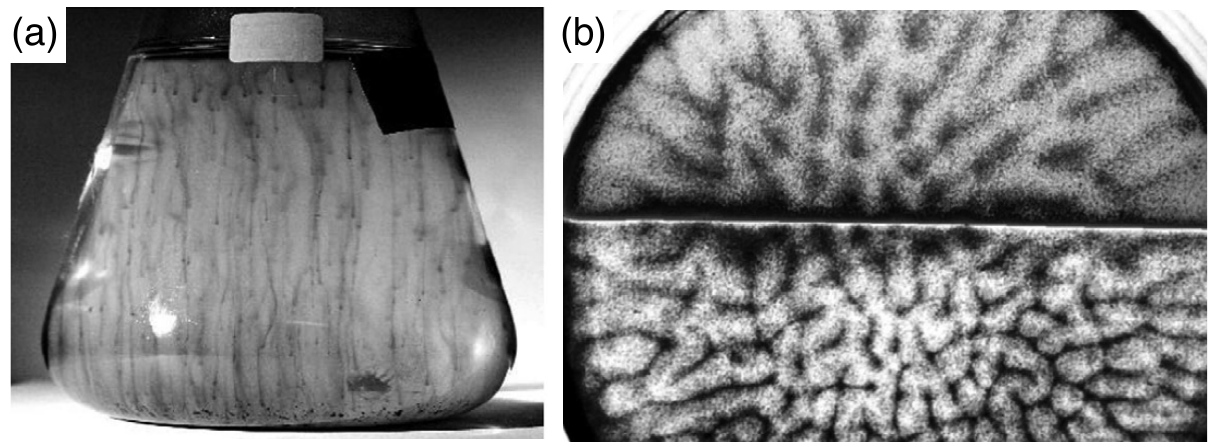}
		\caption{Bioconvective plumes and pattern formation. (a) Long vertical gyrotactic plumes in a culture flask. 
			(b) Pattern formation in
			a shallow Petri dish (depth $0.4\,$cm; width $5\,$cm; $10^6\,$cells/cm$^3$); the illumination is
			white from below, where the lower half is covered with a red filter ($660\,$nm;
			contrast enhanced). The two regions display different bioconvection
			patterns: in the top half, white illumination leads cells to swim upwards,
			with phototaxis supporting gravitaxis and suppressing gyrotaxis, initiating
			an overturning instability with broad downwelling structures; in the bottom
			half, cells do not respond to the red illumination and form finely focused
			gyrotactic plumes. Adapted from \cite{Williams2011a}}
	\label{fig:plumes}
\end{figure}

A quiescent suspension of swimming microorganisms often develops spontaneously large scale currents, whirling the microbes around in mesmerisingly complex flows \cite{Platt1961}. Changing patterns of tight cell-rich downwelling plumes interspersed with large upwells of low cell concentration can be observed in intertidal pools \cite{Bearon2006}, and are a very familiar sight in the laboratory (Fig.~\ref{fig:plumes}). They reveal macroscopically the incessant activity that is present at the microscale, which distinguishes in a fundamental way the behaviour of these so-called active complex fluids \cite{Fodor2018} from their passive counterparts like colloidal suspensions. 
The emergent hydrodynamic instabilities, called bioconvection, are amongst the oldest reported collective effects in microbial suspensions \cite{Wager1911,Loefer1952,Robbins1952,Platt1961,Plesset1974}.
Proposed as potentially important to enhance microbial growth \cite{Bees2014,Kessler1997}, the ecological impact of bioconvective flows has been extensively debated \cite{Janosi2002,Bees2014}. Recently they have been shown to play an important role in promoting growth within stratified lakes \cite{Sommer2017}, where biomixing is able to homogenise chemicals over thicknesses of up to $2\,$m.
Although bioconvection is often a consequence of the natural tendency of microorganisms to swim upwards due to non-uniform mass distribution within their bodies (bottom-heaviness) \cite{Pedley2010c,Pedley1992} , in many cases it results instead from cells accumulating in response to environmental physico-chemical stimuli.
Responses to chemoattractants (chemotaxis) \cite{Tuval2005} and light (phototaxis) \cite{Suematsu2011,Williams2011,Arrieta2019} have both been observed to cause or alter bioconvection in microbial suspensions. In turn, this raises the question of whether the external control of  stimuli could be used to govern large scale hydrodynamic instabilities in suspensions of swimming microorganisms, thereby harnessing the activity of the suspension to govern its macroscopic transport properties.
Control of chemical gradients can be achieved within microfluidic devices \cite{Salek2019,Sung2018,Menolascina2017,Rusconi2014a}, but advection of the chemical species by the underlying fluid \cite{Lushi2012,Lushi2018,Ryan2019} makes this a complex strategy for robust control of flows in macroscopic active matter suspensions.
Conversely, arbitrary spatiotemporal patterns of light can be engineered easily, and are independent of ensuing flows within the suspension. Although light scattering by cells can be a complicating factor \cite{Ghorai2013}, light affords -in principle- an easy route to control microbial motility, either in speed (kinesis) or direction (taxis). This includes naturally phototactic microalgal species \cite{Jekely2009}, some of which are technologically important \cite{Razeghifard2013,Ho2014}, as well as bacteria with genetically engineered light response \cite{Arlt2019,Frangipane2018}. Overall, light is a promising tool to determine microbial accumulation \cite{Giometto2015,Martin2016,Arrieta2017}, and through this modify and control macroscopic flows within active suspensions.

Here we will review current understanding of the interaction between microbial phototaxis and bioconvection (for a recent general  review on bioconvection , see \cite{Bees2020}). After summarising the main mechanism leading to bioconvection in Sec.~\ref{sec:bioconv}, Sec.~\ref{sec:mechano} will discuss the effect of phototactic perturbations to an underlying bioconvective instability. We will then proceed to review  recent studies that are pioneering the use of phototaxis to drive -rather than perturb- hydrodynamic instabilities (Sec.~\ref{sec:photobioconv}), and the combination of phototaxis and externally imposed shear flows to alter the macroscopic distribution of phototactic cells (Sec.~\ref{sec:photofocussing}). We will then conclude in Sec.~\ref{sec:conclusion}.
\label{sec:intro}

\section{Bioconvection of microbial suspensions: a quick primer}
\label{sec:bioconv}
 \begin{figure*}[!t]
 	\begin{center}
 	\includegraphics[width=0.8\linewidth]{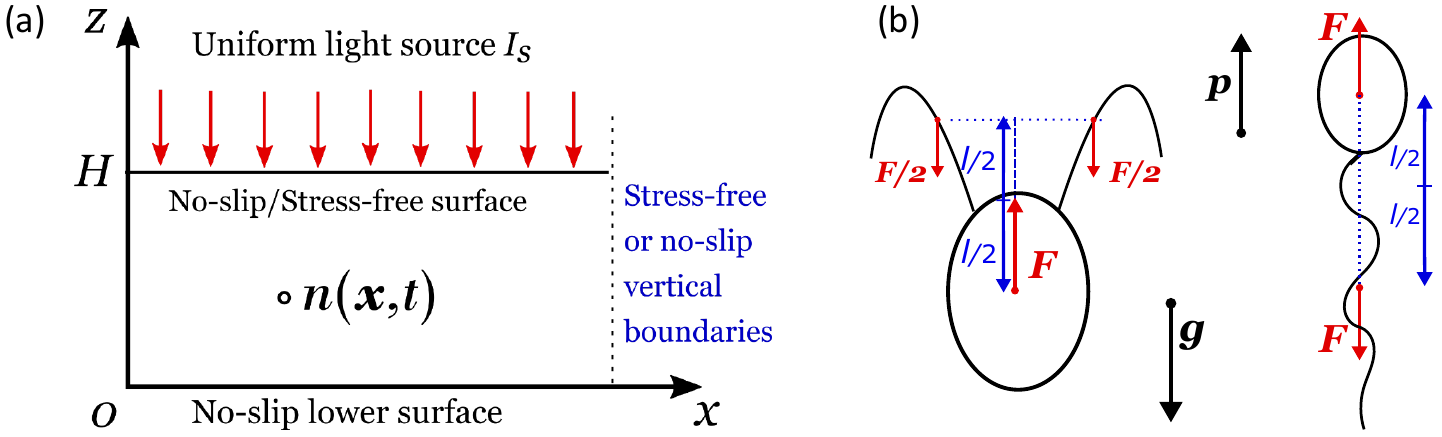}
 		\caption{Standard setting for photo-bioconvection. (a) The cell suspension, of instantaneous local cell density $n(\bm{x},t)$, is contained within a horizontal chamber of thickness $H$. The bottom surface is no-slip, while the top surface can be either no-slip or no-stress. The suspension is subjected to a uniform vertical illumination of intensity $I_s$. (b) Typical structures of the main types of swimming micro-organisms. Left: a biflagellate microalga (puller-like) Right: a bacterium (pusher-like). Both swim along the direction $\bm{p}$ and are subjected to gravity, with gravitational acceleration  $\bm{g}$.}
 		\end{center}
 		\label{fig:schematic}
 \end{figure*} 
Bioconvection, the macroscopic recirculation that develops spontaneously in concentrated suspensions of micro-organisms, was first observed for bottom-heavy microalgae  \cite{Platt1961}. 
In these microorganisms, the centre of mass lies behind the hydrodynamic centre of resistance. This is commonly ascribed to inhomogeneities in the mass distribution within the microbial cell body, although shape asymmetries have been proposed to play an important role \cite{Roberts1970,Roberts2006}. Indeed, these have been shown to be the dominant cause of bottom-heaviness in the model microalga {\it Chlamydomonas reinhardtii} \cite{Kage2020}. The gap between the centres of mass and hydrodynamic resistance results in a gravitational torque which biases  cell motion upwards \cite{Kessler1985}. This phenomenon, called gravitaxis, accumulates bottom-heavy cells at the upper surface. As cells are usually denser than the surrounding fluid, the accumulation makes the ensuing stratification gravitationally unstable, leading to the development of plumes and convective rolls in a manner similar to Rayleigh-B\'enard convection. Sinking bioconvective plumes are then reinforced by gyrotaxis, the tendency of bottom-heavy cells to move towards downwelling regions resulting from the balance between gravitational and hydrodynamic torques originally observed in pipe flow by J. Kessler \cite{Kessler1985}.
A continuum model for gyrotactic bioconvection of a dilute suspension of swimming microorganisms was first studied by Pedley, Hill and Kessler \cite{Pedley1988} and then by Pedley and Kessler \cite{Pedley1990a,Pedley1992} (Fig.~\ref{fig:schematic}a). Here, the suspension of algal cells is modelled as a continuum of identical prolate ellipsoids of volume $v$ and density $\rho_c$, that swim with a constant speed of $V_s$ along a direction $\bm{p}$ and do not interact directly with each other. The swimming direction is determined by a combination of environmental torques acting on the cell and the randomness inherent in microbial swimming, and is therefore described locally by a probability distribution $f(\bm{p})$. This is taken to satisfy a quasi-steady Fokker-Planck (FP) equation, balancing advection by currents in $\bm{p}$ -due to gravitational and hydrodynamic torques- with an effective angular diffusion of diffusivity $D_R$ due to cell swimming.
If the concentration of cells at position $\bm{x}$ and time $t$ is $n(\bm{x},t)$, and $\bm{u}(\bm{x},t)$ and $p$ represent the velocity and pressure of the underlying fluid, the governing equation for the system are \cite{Pedley1990a}
\begin{align}
\label{eq1}
&\nabla\cdot\bm{u} =0,\\
\label{eq2}
&\rho\left( \frac{\partial \bm{u}}{\partial t} + \bm{u}\cdot\nabla \bm{u}\right) = -\nabla p +\nabla\cdot\left(2\mu \bm{E} + \bm{\Sigma}\right)+nv\Delta \rho \bm{g},\\
\label{eq3}
&\frac{\partial n}{\partial t} + \nabla\cdot\left( n\left(\bm{u}+\bm{V}\right)\right) = \nabla\cdot\left(\mathbf{D}\cdot\nabla n\right),\\
\label{eq4}
&\nabla_{\bm{p}}\cdot\left(f \dot{\bm{p}}\right) = D_R \nabla_{\bm{p}}^2 f,\\
\label{eq5}
&\dot{\bm{p}} = (\mathbf{I}-\bm{p p})\cdot\bm{\Omega}\cdot\bm{p} + \Lambda\left(\mathbf{I} - \bm{pp}\right)\cdot\bm{E}\cdot\bm{p} + \frac{1}{B}(\mathbf{I}-\bm{p p})\cdot\bm{k}.
\end{align}
Here we recognise the Navier-Stokes equations for a fluid of density $\rho$ and dynamic viscosity $\mu$. Mass conservation, Eq.~\eqref{eq1}, enforces fluid incompressibility; and momentum balance, Eq.~\eqref{eq2}, is governed by the rate of strain tensor, $\bm{E} =(\nabla\bm{u} + \nabla\bm{u} ^{\mathrm{T}})/2$, the buoyancy forces caused by the excess cell density ($\Delta\rho = \rho_c-\rho$) considered within the Bousinnesq approximation (with $\bm{g}$ being the gravitational acceleration), and the excess deviatoric stress $\bm{\Sigma}$ which will be discussed in detail below.
Local cell conservation, Eq.~\eqref{eq3}, depends on the cells' effective diffusion tensor $\mathbf{D}$, which is non-isotropic for gravitactic species \cite{Hill2002}, and on advection by both the local fluid flow and swimming speed $\bm{V}$. The cells' gravitational sedimentation has been neglected in this formulation, but its effect was later investigated by Pedley \cite{Pedley2010c,Hwang2014}.  
The swimmer velocity $\bm{V} = V_s \left\langle \bm{p}\right\rangle $, where $\left\langle \bm{p}\right\rangle $ is the local average swimming direction: %
\begin{equation}
\left\langle \bm{p}\right\rangle = \int_{S^2} \bm{p} f(\bm{p})d^2 \bm{p}.
\label{eq6}
\end{equation}
Note that, although $\bm{p}$ is a unit vector, the magnitude of $\left\langle \bm{p}\right\rangle $ can range from 0 (completely random swimming) to 1 (all cells swimming along the same direction). 
Changes in the local distribution of cell orientation, and the effect of local environmental torques on the swimming direction $\bm{p}$ are modelled by Eqs.~(\ref{eq4},\ref{eq5}). Here, the distribution function $f(\bm{p})$ is the solution of a quasi-steady advection-diffusion equation on the unit sphere $S^2$ with effective rotational diffusivity $D_R$. This approach slaves $f(\bm{p})$ to the instantaneous local orientational currents $\dot{\bm{p}}$, a valid assumption if temporal changes in the local flow structure are slow compared with the intrinsic equilibration timescale of the local distribution of cell orientations. 
Equation~\eqref{eq5}, in turn, describes how a cell's swimming direction $\bm{p}$ changes due to viscous and gravitational torques. The former is a combination of uniform rotation (from the local vorticity $\bm{\Omega}=(\nabla\bm{u} - \nabla\bm{u} ^{\mathrm{T}})/2$), and alignment along the principal extensional direction (from the rate of strain tensor $\bm{E}$). The relative importance of these terms depends on the shape parameter $\Lambda$, which measures the cell's eccentricity \cite{Pedley1987}. It is zero for a spherical organism, and tends to unity for rod-like swimmers \cite{Pedley1987}. This is responsible for the famous Jeffery orbits of elongated ellipsoids in shear flow \cite{Jeffery1922}.
The last term on the right hand-side, instead, is the gravitational torque due to bottom-heaviness, with the gravitational reorientation timescale $B$ determined by the balance between the maximal gravitational torque and the viscous resistance to rotation \cite{Kessler1988}:
\begin{equation}
B = \frac{\mu \alpha_{\perp}}{\rho_c gh}.
\label{eq7}
\end{equation}
Here $\alpha_{\perp}$ is the resistance coefficient for the cell's rotation about an axis perpendicular to $\bm{p}$, and $h$ is the centre of mass offset of the cell. Note that the physical implication of any term of the form $(\mathbf{I} - \bm{pp})\cdot\bm{e}$ on the right-hand-side of Eq.~\eqref{eq5} is to rotate and align $\bm{p}$ along $\bm{e}$.

We have not yet discussed the term $\bm{\Sigma}$, added to the stress tensor in  Eq.~\eqref{eq2}. This accounts for  the stress contribution of the cells to the ambient fluid flow, and is the sum of two terms, $\bm{\Sigma} = \bm{\Sigma}_p + \bm{\Sigma}_s$. The former is a passive term due to the disturbance flow field caused by the rigid cell bodies ($\bm{\Sigma}_p$), which follows the expression derived by Batchelor for a suspension of identical rigid passive particles \cite{Batchelor1970}.
The latter is an active stress term from cell motility which comes from the stresslet contribution to the forces exerted by a swimming cell on the surrounding fluid. If the magnitude of the drag force from the body on the liquid is $F$, and this is countered -on average- by an equal and opposite thrust from the motile appendages at a distance $l$ from the hydrodynamic centre of the body, the $\bm{\Sigma}_s$ is given by \cite{Kessler1988}: \begin{equation}
\bm{\Sigma}_s= \pm nFl \left( \langle\bm{p p}\rangle  - \frac{1}{3} \mathbf{I} \right),
\label{eq8}
\end{equation}
where $n$ -as above- is the local cell concentration, and the sign choice depends on whether the microswimmers are puller ($+$) or pusher ($-$) (Fig.~\ref{fig:schematic}b).  Pedley and Kessler \cite{Kessler1988,Pedley1990a} showed that this is by far the dominant contribution, and therefore only the active stress contributions ($\bm{\Sigma}_s$) are taken into account in Eq.~\eqref{eq2}. 
Taking $\bm{\Sigma}\simeq\bm{\Sigma}_s$ and using Eq.~\eqref{eq8} gives a good approximation of the effects of swimmers on the flow in dilute suspensions (with typical cell densities of the order $10^6 \mathrm{cells/cm^3}$) and has been used extensively throughout the literature on collective motion. However, it can also be neglected in many situations with pre-existing background flows that have characteristic viscous stresses much larger than $\|\bm{\Sigma}_s\|$.

Besides contributing to the fluid's stress tensor, cell swimming leads also to an effective diffusivity tensor $\mathbf{D}$. For the constant swimming speed $V_s$ assumed here, it can be derived from the Green-Kubo relations as \cite{Pedley1990a,Sharma2016,Dalcengio2019}
\begin{equation}
\mathbf{D} = \int_{0}^{\infty} \left\langle \bm{V}_r(t) \bm{V}_r(t-t') \right\rangle dt'; \:\:\;
\bm{V}_r \equiv V_s (\bm{p} - \langle \bm{p} \rangle),
\label{DFP}
\end{equation}
which, under the assumption of a constant correlation time $\tau$ reduces to \cite{Pedley2010c}
\begin{equation}
\mathbf{D} = V_s^2 \tau \left\langle (\bm{p} - \langle \bm{p}\rangle)(\bm{p} - \langle \bm{p}\rangle)  \right\rangle.
\label{DFP2}
\end{equation}
Typical values of $\tau$ range between $1\,$s and $5\,$s.
These expressions assume that the background fluid shear does not influence fluctuations and persistence in cell orientation.  
The validity of this approximation in a real system is uncertain, especially for large shear rates. As a result, Bees and Hill (for spherical swimmers \cite{Hill2002}) and Manela and Frankel (for axisymmetric swimmers \cite{Manela2003}) have used instead a generalised Taylor dispersion theory (GTD) to formulate the diffusivity of a suspension of gyrotactic micro-swimmers. 
For sufficiently low shear rates the difference between the two approaches is not significant, and the previous -simpler- approach to $\mathbf{D}$ can be used for example to study the initial development of bioconvection from a quiescent background fluid. 
The stability of the suspension is generally described in terms of a Rayleigh number for active particles $R=\bar{n}gv\Delta\rho H^3/\left(\mu D\right)$, which depends on the ratio between buoyancy and viscous forces  and the ratio between diffusion of momentum within the fluid and active diffusion. Here $\bar{n}$ is the average cell concentration and $H$ is the container thickness.
Through linear stability analysis of Eqs.~(\ref{eq1}-\ref{eq5}), it is possible to predict the onset of the instability, in terms of a critical Rayleigh number; and the most unstable wavelength, usually called the bioconvective wavelength, which should then correspond to the characteristic separation between plumes observed experimentally (Fig.~\ref{fig:ghorai05}b,c) \cite{Hill2005,Bees2020}.

The approach that we have briefly outlined offers in general a good qualitative description of the experimental observations of bioconvection. Quantitatively, however, the agreement is often less good. For shallow suspensions, for example, theoretical predictions overestimate the observed wavelengths (observed: $\sim1-3\,$mm, vs. predicted: $\sim4-7\,$mm \cite{Kessler1985b,Pedley1988,Bees1997}), although better agreement can be obtained by leaving constitutive microbial parameters ($B,\tau,V_s\dots$) as variables to be fitted \cite{Bees1998}.
Pedley and Kessler have also suggested that the actual non-linear nature of the instability might be a cause of the observed discrepancy between the characteristic wavelength for bioconvection observed experimentally, and the one predicted from linear stability analysis \cite{Pedley1990a}. Much of the quantitative uncertainty must come also from our limited knowledge of the swimming behaviour of the microorganisms within the suspension, and possibly of their ability to sense and react to mechanical stresses \cite{Rafai2010,Lele2013}. 
Finally, bioconvection experiments show clearly three dimensional patters. Therefore, in principle, the full three-dimensional problem should be solved for a true quantitative comparison between theory and experiments. Recent 3D simulations are starting to address this issue \cite{Ghorai2007,Karimi2013,Ghorai2015}, and in particular Ghorai et al. (2015) \cite{Ghorai2015} found characteristic wavelengths that were in good agreement with previous experiments\cite{Bees1997}, provided that the diffusion due to randomness in cell swimming behaviour is small. 

\section{Phototactic modifications to standard Bioconvection}
\label{sec:mechano}
\begin{figure*}[!t]
 	\begin{center}
 		\includegraphics[width=0.8\textwidth]{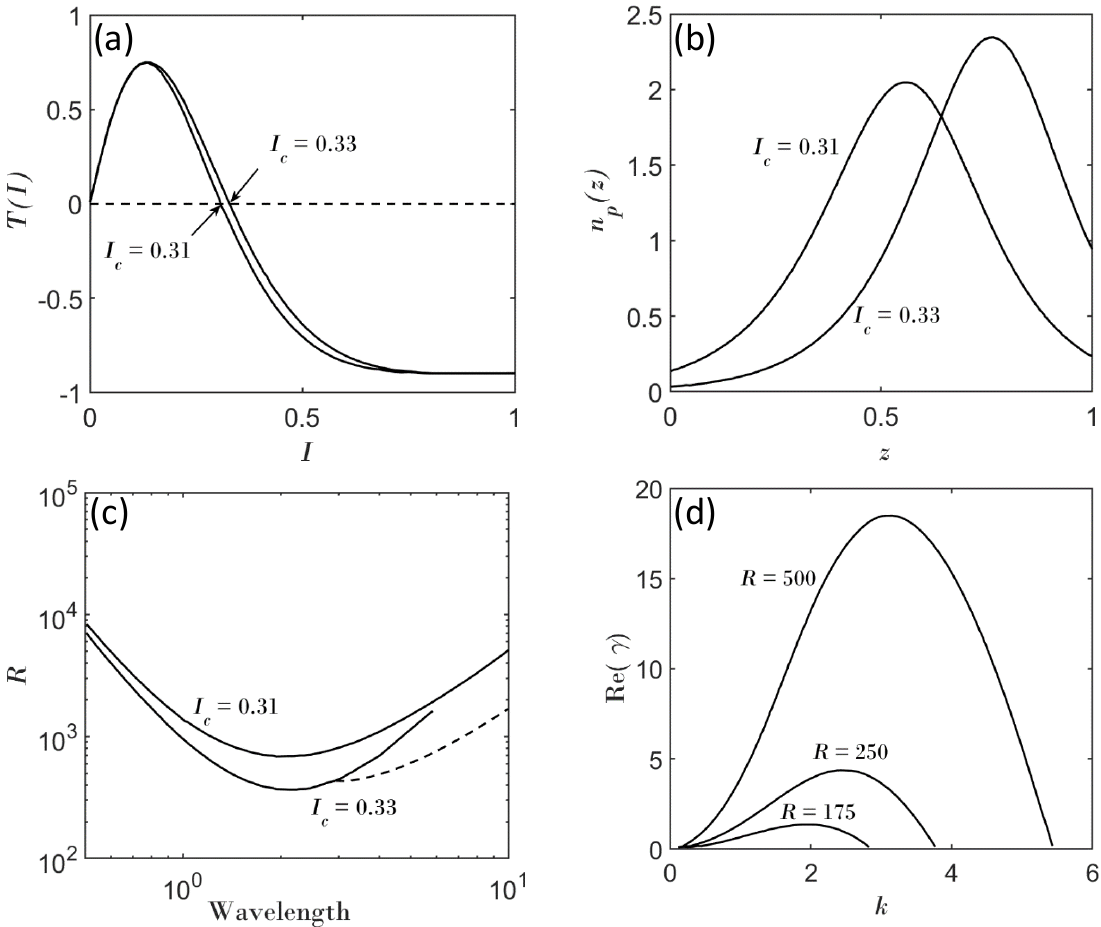}
 		\caption{Penetrative bioconvection. (a) Taxis functions corresponding to different values of $I_c$ (non-dimensionalised), and (b) corresponding equilibrium profiles of cell concentration (for $I_s ~= ~0.5$). In this example, despite being close in value, the critical light intensities are well separated in space due to the slow decay of the intensity profile ($\alpha\bar{n}H=1$). (c) Neutral curves separating stable perturbations (below) from unstable ones (above). (d) The perturbations' growth rate depends strongly on the Rayleigh number. Adapted from  \cite{Ghorai2005,Ghorai2010}.}
 	\end{center}
 	\label{fig:ghorai05}
\end{figure*}

The general framework presented in Sec.~\ref{sec:bioconv} served as a starting point for the first studies on the effects on bioconvection of phototaxis,  the ability of some microbial species (e.g. green microalgae \cite{Foster1980} and dinoflagellates \cite{Moldrup2012}) to swim towards a light source (positive phototaxis) or away from it (negative phototaxis). The light needs to be within the range of wavelengths which the light-sensitive organelles are sensitive to (between $\sim420-500\,$nm for {\it Chlamydomonas} \cite{Crescitelli1992}), which overlaps with -but does not necessarily exactly correspond to- the photosynthetically active range \cite{Harris2009}.
The pioneering continuum studies of photo-bioconvection \cite{Yamamoto1992}, considered purely phototactic cell suspensions that are exposed to uniform illumination from above or below (Fig.~\ref{fig:schematic}a). Swimming cells are still assumed to be slightly heavier than the ambient fluid, but gyrotaxis and gravitaxis are neglected and the up-swimming is exclusively due to phototaxis. 
The swimmer velocity, however, depends on the intensity of light $I$, with positive phototaxis for light intensity below a critical value $I_c$ ($I<I_c$), and negative phototaxis otherwise ($I>I_c$). This captures the fact that phototactic organisms move away from intense light to avoid photodamage \cite{Jekely2009}, and results in a tendency to accumulate cells where $I\approx I_c$. The swimming velocity in Eq.~\eqref{eq3} is then 
\begin{equation}
\label{photoV}
\bm{V} = V_s T(I) \bm{k} , \;\;\;\;\mathrm{where}\;\;\;\;
T(I)\begin{cases}
\geq 0, \; \; \; \;  \; \; \; \; I\leq I_c\\
< 0, \; \; \; \;  \; \; \; \;  I > I_c
\end{cases}.
\end{equation} 
Here $\bm{k}$ is the vertical unit vector and $T(I)$ is the phenomenological taxis function with values between -1 and +1 (Fig.~\ref{fig:ghorai05}a).  The shape of taxis function is -in principle- species-dependent, and is usually constructed based on the simplification of controlled experimental observations \cite{Ghorai2005}. At the same time, the diffusion tensor is assumed to be constant and isotropic. This non-monotonic response to light is coupled to changes in light intensity due to the absorption and scattering of light by cells, or ``shading''. This phenomenon is particularly important for dense cell suspensions like those that can be achieved in the laboratory or within bioreactors. For thick but sufficiently dilute suspensions, the local scattering of light by the cells is weak and the light intensity $I(\bm{x},t)$ is simply given by the Beer-Lambert law \cite{Ghorai2010}: 
\begin{equation}
I(\bm{x},t) =  I_s \,\text{exp}\left( -\alpha \int_{z}^{H} n(\bm{x},t) dz \right),
\label{Ix}
\end{equation} 
where $I_s$ is the uniform light intensity at the upper surface of the suspension ($z=H$), and $\alpha$ is the absorption coefficient. 
The rigid bottom with no slip boundary condition is located at $z=0$
Light intensity is therefore a monotonically decreasing function of depth ($z$).
If $I_c$ lie between the maximum and minimum intensities across the suspension, a sublayer of cell concentration is formed in the interior of the chamber (Fig.~\ref{fig:ghorai05}b). The region above this sublayer is gravitationally stable and it is resting on a gravitationally unstable region below. The bottom layer can then develop circulating flows which penetrate the stable layer, resulting in bulk motion throughout the suspension \cite{Vincent1996}. This phenomenon is called penetrative bioconvection. 
Linear instability analysis showed that, within this model, the onset of the instability and the length scales of the ensuing patterns depend on the average cell concentration, the position at which $I=I_c$ for a uniform suspension, and the depths of the gravitationally stable and unstable layers  (Fig.~\ref{fig:ghorai05}b,c). 
Depending on the values of the parameters, this model can develop two different types of large scale flows, either steady or periodic. 
The former is the usual type of  bioconvective flow. The latter corresponds to oscillating flow and concentration fields that  occur when upswimming due to phototaxis is dominant over convective downwelling. Panda and Singh \cite{Panda2016} extended Ghorai and Hill's work to show that these instabilities are opposed by the presence of rigid side walls, which enhance stability for some parameter regimes.

The models above usually rest on two major assumptions: they neglect light scattering from the cells, implying that each cell only receives light vertically from the uniformly illuminating source; and they neglect bottom-heaviness and therefore gyrotaxis. 
Let us consider these assumptions individually. If light scattering from cells is important, then each cell will have to respond to multiple stimuli besides the main one from the illuminating source.  This was first discussed by Ghorai {\it et al.} \cite{Ghorai2010} by introducing a radiative transfer equation for the direction-dependent light intensity, and a scattering probability that determines the likelihood for light to be scattered by a given amount. Both isotropic and anisotropic scattering probabilities were  studied \cite{Ghorai2013,Panda2013}.
The inclusion of scattering complicates the model significantly, and makes it a formidable task to find the intensity profile $I(\bm{x})$, which  needs to be solved for together with the fluid flow and cell concentration. Linear stability analysis \cite{Ghorai2010} and 2D numerical solutions \cite{Panda2013} reveal that for purely scattering suspensions, the intensity $I(\bm{x})$ may not vary monotonically with depth. A suspension can therefore exhibit more than one location where $I = I_c$, leading to an altered base state from the single sublayer discussed earlier. Besides this difference, light scattering from cells does not seem to introduce any qualitatively new instabilities within the suspension \cite{Ghorai2010}. However, it should  be stressed that these results hinge on a specific simple model of phototaxis. Light scattering by cells generates  light fields with multiple sources, and our knowledge of phototaxis in this situation is currently poor.  Although it has been investigated for gliding motility in cyanobacteria \cite{Yang2018}, it is not a well-understood process for swimming cells.

Thus far, we have only considered models for pure phototaxis, where upswimming was just a consequence of cells' motion towards light. Many microorganisms, however, are also gravitactic and gyrotactic and should -in principle- be taken into account to describe the suspension's dynamics. 
The first model of this kind, which modified Pedley and Kessler's model (Sec.~\ref{sec:bioconv}) to account for phototaxis, was proposed by Williams and Bees \cite{Williams2011}. They considered three different models to incorporate the effects of phototaxis: a photokinesis model where cell's swimming speed is a function of light intensity (A); a coupled photo-gyrotactic model where light controls the position of a cell's centre of mass, and therefore its bottom-heaviness (B); and an extra ``phototactic torque'' in Eq.~\eqref{eq5} to reorient cells along gradients of light intensity (C).
We note that it has recently been demonstrated that some microbial species can modify their fore-aft mass asymmetry in response to mechanical stimuli \cite{Sengupta2017}, and it would not be far fetched to expect that, in some cases, this might also be true for light stimuli.
Overall, the models proposed by Williams and Bees are an excellent starting point to investigate the coupling between photo- and gyro-taxis. They lead to changes in Eqs.~(\ref{eq3},\ref{eq5}), which become 
\begin{align}
\label{dndtI}
&\frac{\partial n}{\partial t} + \nabla\cdot n\bigg(\bm{u} + \underbrace{V_s \left(1-\frac{I}{I_c}\right)}_\text{(A)} \langle\bm{p}\rangle \bigg) = 
\nabla\cdot(\bm{D}\cdot\nabla n) \\
\label{pdotI}
&\dot{\bm{p}} = \underbrace{\frac{1}{B(I)}}_\text{(B)}(\bm{I - pp})\cdot\underbrace{\bm{k}(I)}_\text{(C)}+(\bm{I - pp})\cdot(\bm{\Omega}\cdot\bm{p}+\Lambda\bm{E}\cdot\bm{p}),
\end{align}
where $V_s$ is now the swimming speed in the absence of any illumination, and $I_c$ is the critical intensity from Eq.~\eqref{photoV}.
Model (A) includes an intensity-dependent speed in Eq.~\eqref{dndtI} which encodes both a photokinetic effect (i.e. swimming speed that depends on light intensity) as well as the ability of the cells to reverse their swimming direction in regions where $I>I_c$. 
The other two models, instead, modify the equation for the cell orientation (Eq.~\eqref{eq5}), and therefore the value of $\langle\bm{p}\rangle$. The new equation, Eq.~\eqref{pdotI}, includes either an intensity-dependent gravitational reorientation timescale $B(I)$ (model B) or an intensity dependent preferred cell orientation $\bm{k}(I)$ (model C).
The former is an ad hoc manifestation of the cell's ability to change its centre of mass offset as a function of $I$, which is taken to follow
\begin{equation}
B(I) = \frac{\mu \alpha_\perp}{\rho_c g h_0 (1-I/I_c)},
\label{BI}
\end{equation}
where $h_0$ is the centre of mass offset in the dark (see also Eq.~\eqref{eq7}). This assumes that bottom-heavy cells become top-heavy for $I>I_c$.
 The latter is the effect of the phototactic torque. The preferred cell orientation $\bm{k}(I)$ is obtained by a balancing the gravitational torque $\bm{L}_g = (\rho_c vgh_0)\bm{p} \times \bm{k}$, and the phototactic torque
\begin{equation}
\bm{L}_p = -\frac{4f_m}{I_c^2 }I(I-I_c) \bm{p} \times (\beta_1\bm{\pi} + \beta_2 \nabla I).
\label{Lp}
\end{equation}
Here $f_m$ is a parameter defining the overall strength of the phototactic torque, while $\beta_1$ and $\beta_2$ quantify the cell's phototactic response to light incident from an arbitrary direction $\bm{\pi}$, and to local gradients in the light intensity. Then  $\bm{k}(I)$ is simply given by 
\begin{equation}
\label{kI}
\bm{k}(I) = \bm{k} +\frac{4f_m}{mgh_0}(I/I_c)(I/I_c-1)(\beta_1\bm{\pi} + \beta_2 \nabla I).
\end{equation}
Williams and Bees studied these three models separately, with two cases considered for model C: $(\beta_1,\beta_2)=(1,0)$ (C\,I), or  $(0,1)$ (C\,II). In general, phototaxis will affect the probability distribution $f(\bm{p})$ (Eq.~\eqref{eq4}), and through this the quantities $\langle \bm{p} \rangle$ (Eq.~\eqref{eq6}) and $\bm{D}$ (Eq.~\eqref{DFP}) in a non-trivial way \cite{Williams2011}. 
The coupled photo-gyrotaxis models are able to predict both steady and oscillating flow patterns. Interestingly models A, B and C\,I produce very similar results, when illuminated from above, which also agree qualitatively with experiments carried out by Williams and Bees \cite{Williams2011a}.
 The basis for this comparison with experiments, however, was deemed questionable for several reasons. Firstly, the experimental methods did not allow for the theoretical base state to be reached before instabilities appeared; secondly there was a lack of reliable direct estimates of the critical intensity $I_c$ for the suspension; and finally it was unclear whether the wavelengths being measured in the experiment corresponded to the
critical state obtained using the theory. Nevertheless, the qualitative agreement observed between models and experiment is promising.
At the same time, model C\,II differs significantly  from the others due to the existence of non-hydrodynamic instabilities. In this case, phototactic torques due to $\nabla I$ lead to horizontal modulations of the cell concentration profile, with equally spaced clusters of cells just above the depth at which $I=I_c$ for the (unstable) base state. This modulation should happen without generating a background fluid flow. 
Finally, we note that a formal extension of generalised Taylor dispersion (GTD) to a suspension of phototactic microorganisms, in the spirit of the work of Hill and Bees \cite{Hill2002}, and Manela and Frankel (for gyrotaxis) \cite{Manela2003}, and Bearon (for chemotaxis) \cite{Bearon2011,Bearon2012,Bearon2015}, has not been attempted yet. It will be interesting to see how GTD alters the stability dynamics of phototactic suspensions.

\section{Bioconvection driven by Phototaxis} 
\label{sec:photobioconv}

A recent experimental endeavour is to use light as the main factor leading to bioconvection. In these studies, no bioconvection patterns are observed in absence of light. Convective instabilities only appear in presence of specific illumination patterns, used to generate the cell-density inhomogeneity that drives bioconvection.

The first instabilities of this type were investigated in suspensions of the microalga \textit{Euglena gracilis} \cite{Suematsu2011,Suematsu2014}. Initially homogenous cultures, filled the gap within a $2\,$mm-thick horizontal Hele-Shaw cell, and were then exposed to strong uniform illumination from the bottom to induce negative phototaxis. 
Within a few minutes, cells accumulated in randomly distributed high-density clusters \cite{Suematsu2011}, which then gathered towards the centre of the chamber. These localised patterns disappeared in absence of light.
The number of clusters and the mean separation between neighbouring clusters can be used to characterise the dynamics of the system as a function of  depth of the chamber and average cell concentration. Both quantities showed ageing, with the former increasing and the latter decreasing as a function of time.  An effective model based on transition probabilities between cell layers was able to recapitulate the qualitative dynamics of cell concentration observed experimentally, including the process of pattern formation, cell circulation around a cluster, and dependence of number of clusters and their distance to nearest neighbours on depth and cell density. This agreement provided support for the hypothesis that pattern formation is a consequence of cell motion transversal to the direction of incoming light, possibly a result of phototaxis to light scattered by the microorganisms themselves. 
Although this model lacks a connection to the underlying hydrodynamics of the suspension, it is still a first step towards understanding the emergence of photo-bioconvective patterns.
A later study of localized photo-bioconvection patterns in the same system \cite{Shoji2014}, showed that the emergence of these high-density cell clusters, or ``bioconvection units'', was dependent on the spatial distribution of the cells before light exposure. Below a critical cell concentration, no bioconvection units were observed for a uniform initial distribution, whereas a single cluster formed for sufficiently spatially heterogeneous suspensions.
This observation suggests that the system is multi-stable, possibly as a result of the topology of the containing chamber.

\begin{figure*}[!t]
	\begin{center}
		\includegraphics[width=0.8\textwidth]{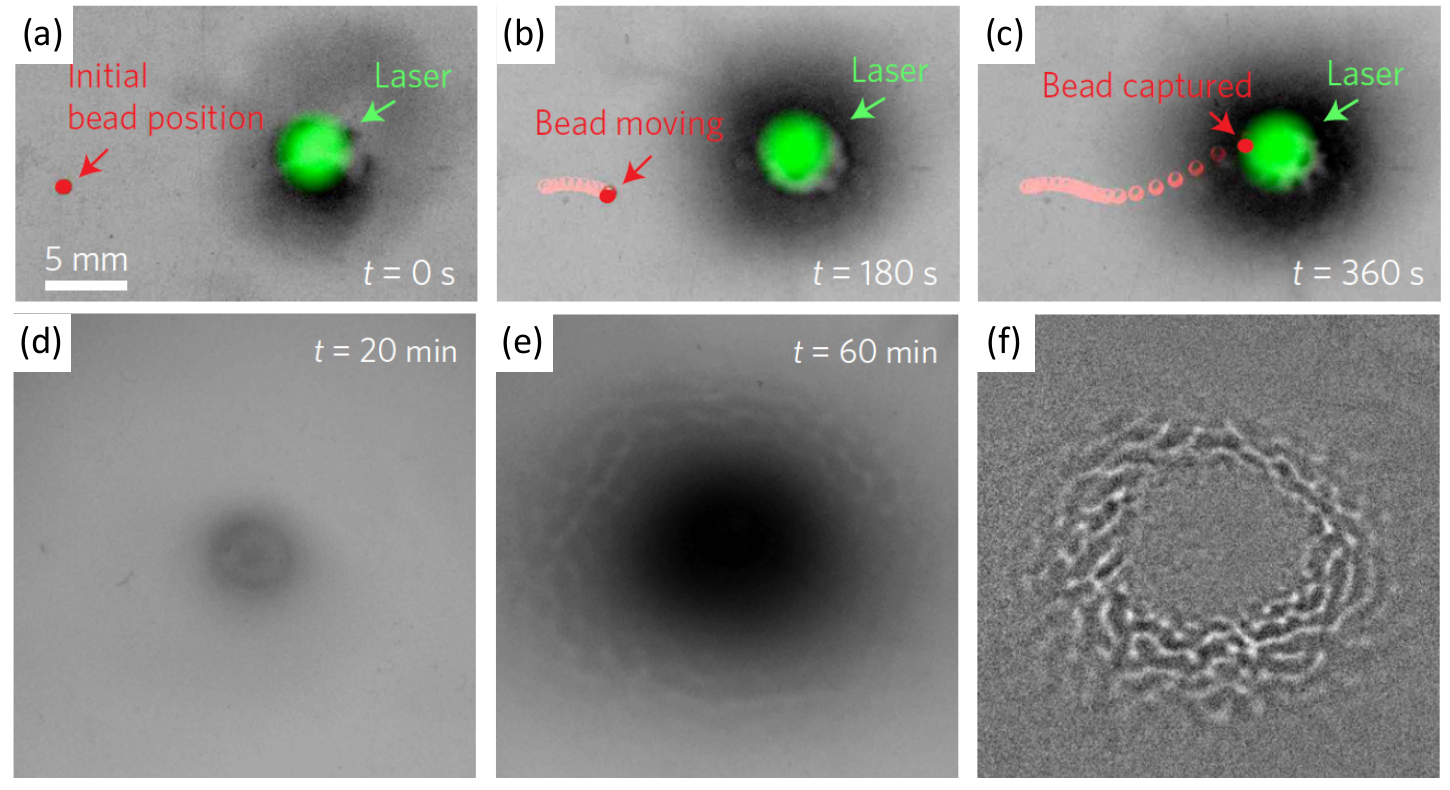}
		\caption{Photo-bioconvection induced by localised laser illumination in \cite{Dervaux2017}. (a-c) An $800\,\mu$m-diameter floating glass bead is transported by the flow field generated by the concentrated cells. (d-f) Time evolution of phototactic cell accumulation at the bottom of the container, and formation of radial waves of concentration. Adapted from \cite{Dervaux2017}}
	\end{center}
	\label{fig:dervaux}
\end{figure*}
In an interesting study by Dervaux {\it et al.} \cite{Dervaux2017}, intriguing bioconvective instabilities were observed instead when a thin suspension of the model unicellular green algae {\it Chlamydomonas reinhardtii} was placed in a Petri dish and illuminated from above by a localised circular laser beam. For intensities within the range of positive phototaxis, the cells accumulated beneath the light beam, and radially symmetric convective flows developed as a result of light-induced collective motion. Although the timescale for the establishment of this instability was long ($\sim1\,$h), the ensuing convergent flows on the surface could be used to collect a floating cargo (here a $800\,\mu$m glass bead) and trap it just above the laser spot (Fig.~\ref{fig:dervaux}a-c). Above a critical Rayleigh number, corresponding to high cell density and layer depth, a novel instability in the form of travelling concentration waves was discovered (Fig \ref{fig:dervaux}d-f). Its development was successfully described through a model for cell reorientation that included turning due to the fluid's vorticity, and a phototactic reorientation similar to the first term in Eq.~\eqref{pdotI}, but with $\bm{k}$ replaced by the direction of $\nabla I$ from the laser. The effective timescale for phototactic turning, equivalent to $B(I)$ in Eq.~\eqref{pdotI}, was left as a fitting parameter. Best fits were obtained for a timescale $\simeq1.2\,$s, significantly faster than the gyrotactic reorientation timescale ($\sim3-6\,$s for {\it Chlamydomonas augustae}, a species known to be strongly gyrotactic \cite{Williams2011a}).
This work is the first experimental validation of the phototactic torque model in a simplified case.

More recently, Arrieta {\it et al.} \cite{Arrieta2019} reported a new example of bioconvection driven by phototaxis, which develops quickly ($\sim30\,$s), and can be easily reconfigured. In this case, a suspension of {\it C. reinhardtii} is loaded within a thin and wide square chamber held vertically, a configuration notably distinct from similar studies in bioconvection, where the widest sides of the chamber are horizontal.
Localised illumination was provided by a $200\,\mu$m-diameter horizontal optical fibre.
With no light from the fibre, cells were uniformly distributed within the suspension. However, as the light was switched on, the algae accumulated phototactically around the fibre, leading to a gravitational instability and the formation of a single localised sinking bioconvective plume. 
 The system was described with a purely phototactic model of 2D bioconvection, based on the Eqs.~(\ref{eq1})-(\ref{eq3}), with no slip at the boundary and no cell flux through the boundary. Following previous studies, \cite{Arrieta2017} the phototactic velocity was taken to be proportional to the gradient in light intensity. The proportionality factor depends on the phototactic sensitivity parameter, $\beta$, previously shown to exhibit an interesting adaptation possibly as a result of the photosynthetic activity of the cells \cite{Arrieta2017}. 
 \begin{figure*}[t]
	\begin{center}
		\includegraphics[width=0.95\textwidth]{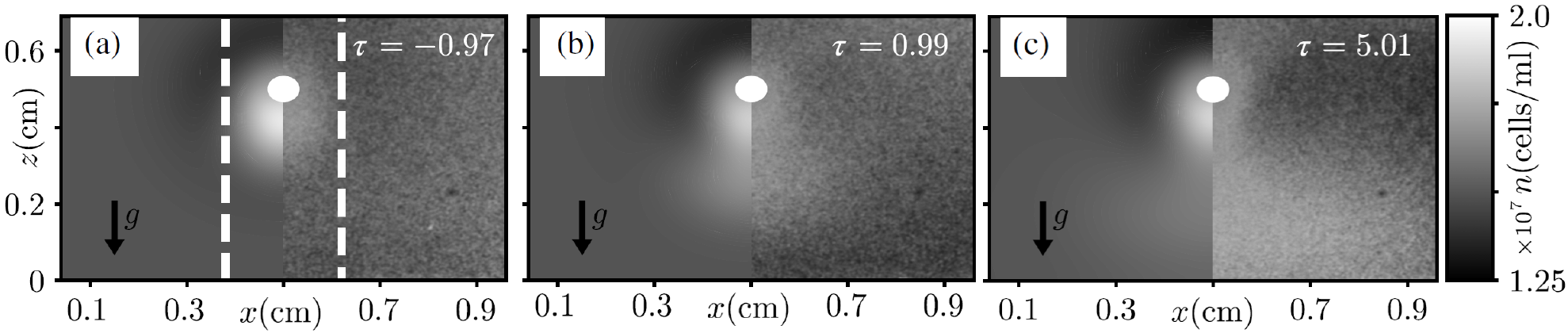}
		\caption{Dynamics of plume formation in localised photo-bioconvection. Cells accumulate around the optical fibre (white disk) and form a single bioconvective plume as the reduced time $\tau$ progresses. Panels show the cell density ($n$) from the continuum model  (left side), and the experiments. Plume formation takes $\sim30\,$s. Adapted from \cite{Arrieta2019}}
 \label{fig:arrieta}
	\end{center}
\end{figure*}
Despite neglecting gyrotaxis and gravitaxis, as well as cell-cell interactions reinforcing alignment at high concentrations \cite{Furlan2012}, the model showed an excellent agreement with experimental observations and was able to capture the structure of the bioconvective flow of cells (Fig.~\ref{fig:arrieta}). It also predicted that for average cell densities below a critical value, $n_c$ ($\sim 10^7 \mathrm{cells/cm^3}$ for $\beta = 0.14$), phototactic accumulation around the fibre would not lead to the development of a plume.
We note, however, that the background fluid has a global recirculation even when the plume does not form, as a result of the excess concentration of phototactic cells around the optical fibre.
 The existence of this bifurcation was confirmed experimentally, and the critical cell density observed was in good quantitative agreement with the one predicted by the model. Overall, the model predicted a bifurcation boundary compatible with the simple relation $\beta n_c = \text{const}$.  Leveraging the speed at which plumes form in this system, the authors then provided a proof of principle that photo-bioconvection can be used to mix the suspension. This was achieved here by alternatively blinking a pair of optical fibres to create two sets of convective flows with crossing streamlines.

\section{Photo-focussing}
\label{sec:photofocussing}
Light can also be used together with externally imposed flows to control the motion of phototactic cells through a mechanism based on balance between hydrodynamic and phototactic torques, analogous to gyrotaxis.  This has been proposed to play a role in the cell accumulation observed by Dervaux {\it et al.} \cite{Dervaux2017}, and can be employed to accumulate cells in specific regions by imposing background flows externally.
This type of cell accumulation, called photo-focussing, has been investigated in a number of recent studies \cite{Garcia2013,Jibuti2014,Martin2016,Clarke2018}. 
\begin{figure*}[t]
	\begin{center}
		\includegraphics[width=0.8\textwidth]{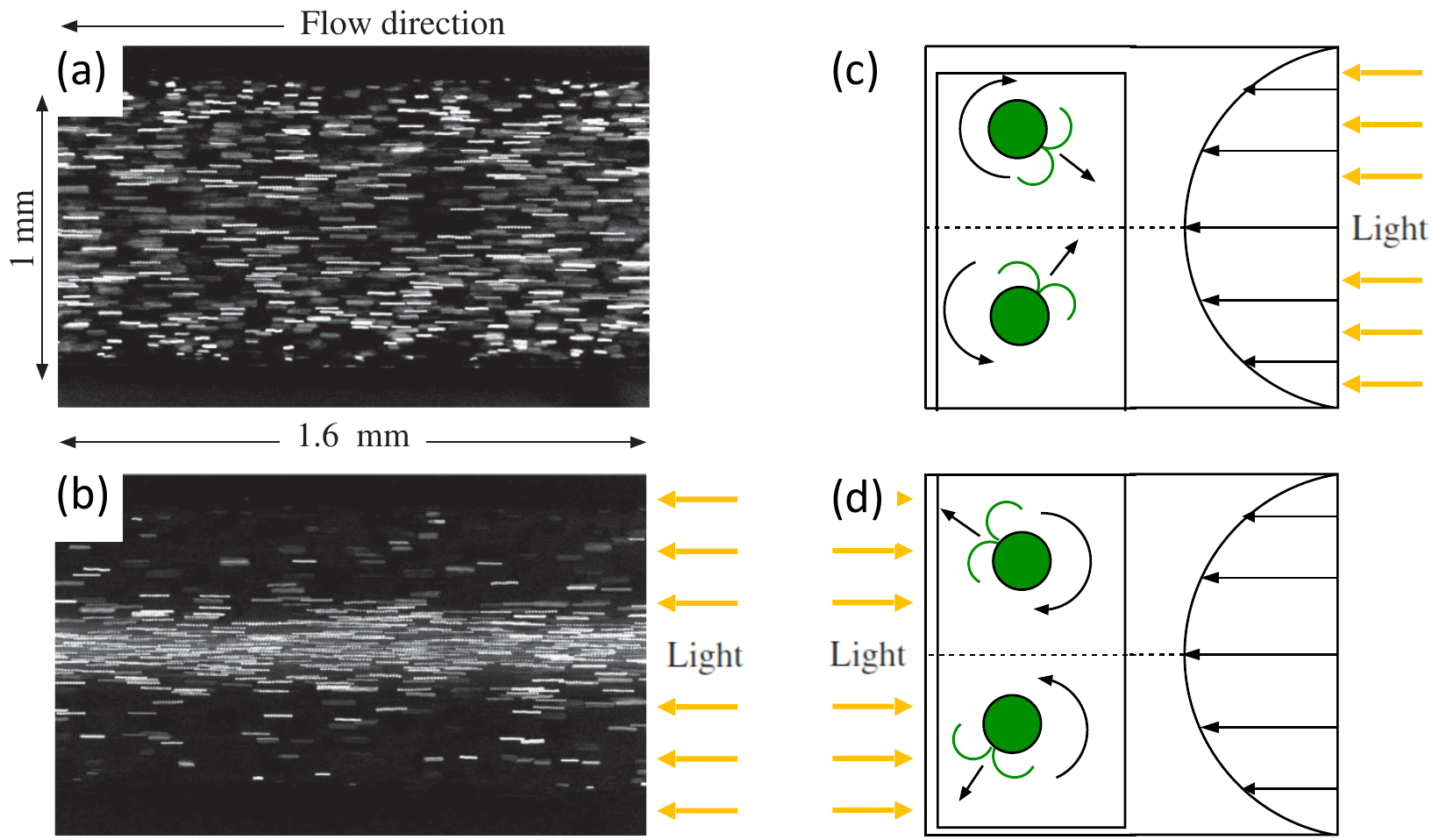}
		\caption{Photo-focussing. Individual trajectories of {\it C. reinhardtii} flowed within a rectangular tube (a) without and (b) with phototactic illumination (light source from the right). (c,d) Schematic of the combined effect of phototaxis and fluid vorticity on the orientation of cells, leading to photo-focussing in the centre (c) or on the sides (d) of the channel. Adapted from \cite{Garcia2013}.}
	\label{fig:salima}
	\end{center}
\end{figure*}
The first direct proof of photo-focussing used a suspension of {\it C. reinhardtii} flowed through a long square capillary of side $1\,$mm \cite{Garcia2013}. The cells' distribution across the channel, which was uniform without light stimuli, could be modulated dramatically by turning on a light source oriented along the length of the channel. When cells tried to phototax downstream, they self-focussed along the axis of the channel; whereas for upstream phototaxis, cells accumulated at the boundaries (Fig.~\ref{fig:salima}). An initial model for photo-focussing used a deterministic description of microbial swimming within a Poiseuille flow \cite{Zottl2012}, modified by an effective phototactic torque on the cell, and provided a good qualitative description of the phenomenon, including the existence of a threshold fluid vorticity beyond which the suspension does not self-focus.
The same system was later studied numerically for a whole suspension of interacting microswimmers \cite{Jibuti2014}, where phototaxis was modelled by the {\it ad hoc} reorientation of each swimmer along the light direction every $\sim1\,$s. The simulations reproduced the formation of a self-focussing jet for sufficiently low vorticities,  and predicted the further fragmentation of these jets into clusters. The latter instability, due to  hydrodynamic interactions between the puller microorganisms within the model, has yet to be observed experimentally in photo-focused jets, although it bears a striking resemblance to instabilities that develop in gyrotactically-focussed cells \cite{Croze2017}.
More recently, the same run-and-tumble-like dynamics was studied in a two-dimensional model of photo-focussing for dilute suspensions, and shown to reproduce well the experimental data for the scaling of the width of the focussing region and the establishment length as a function of the flow velocity \cite{Martin2016}. The agreement between model and experiments provides support for the assumption of an active resistance by the cells to shear-induced turning \cite{Rafai2010}, which was included here as a scaling of the local fluid vorticity by a factor $\eta<1$. Best fits were obtained for $\eta=0.25$. Although the details of the phototaxis model used here are not correct, this approach has the advantage that the steady state concentration profile can be derived analytically. 

\begin{figure*}[!t]
	\begin{center}
		\includegraphics[width=0.95\textwidth]{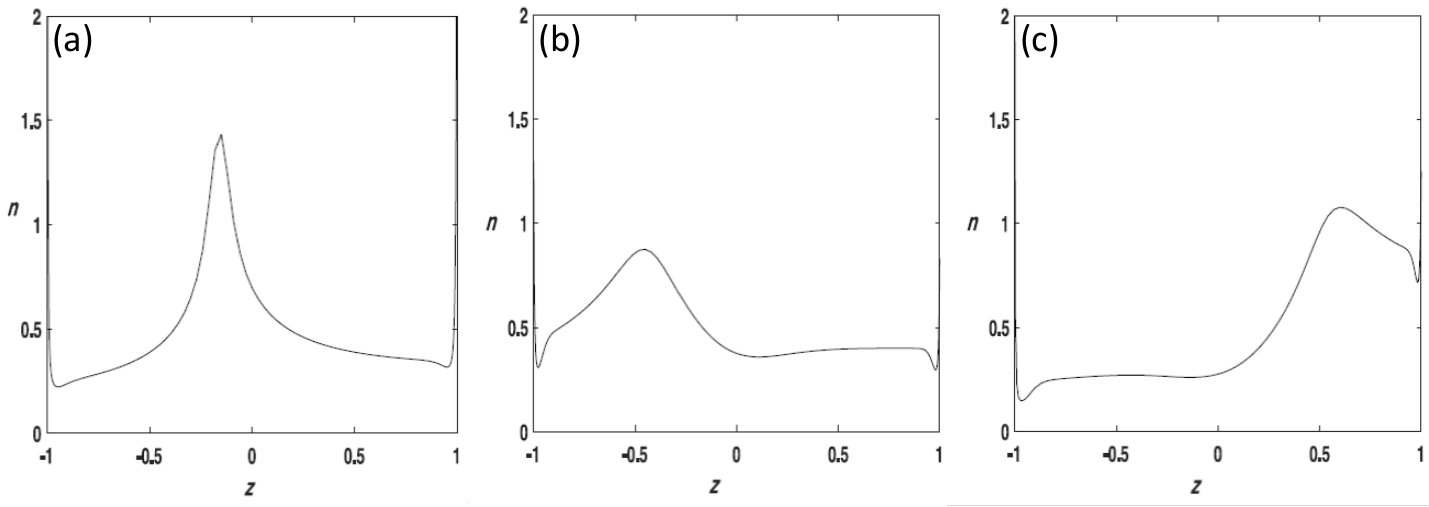}
		\caption{Predicted concentration profiles for phototactic microoganisms within a horizontal Hele-Shaw cell, in presence of a horizontal background flow and vertical illumination. (a) Weak flow ($Pe=1$), strong absorption and $I_s/I_c = 2$. (b) Strong flow ($Pe = 5$), strong absorption, and $I_s/I_c =2$. (c) Strong flow ($Pe = 5$), strong absorption, and $I_s/I_c = 3$. Adapted from \cite{Clarke2018}.}
	\label{fig:clarke}
	\end{center}
\end{figure*}
A systematic theoretical study of photo-focussing of microswimmers in a Poiseuille flow was done by Clarke \cite{Clarke2018}. He considered a suspension of phototactic cells confined between two no-slip horizontal surfaces located at $z=\pm H/2$, and subjected to a background fluid velocity profile $\bm{u} =U(1-4z^2/H^2)\bm{e}_x$.  Unlike the previous studies, the suspension was uniformly illuminated from below. The local distribution function of cell directions,  $f(\bm{x},\bm{p})$, satisfied a generalised form of Eq.~\eqref{eq4}:
\begin{equation}
\label{Clarke1}
(\bm{u} + \bm{V})\cdot\nabla_{\bm{x}}f+ \nabla_{\bm{p}}\cdot(f \dot{\bm{p}}) = D_R \nabla_{\bm{p}}^2 f + D\nabla_{\bm{x}}^2 f ,
\end{equation} 
where $D$ is the translational diffusivity, and $\bm{p}$ satisfies Eq.~\eqref{eq5} with $1/B = 0$.
Phototaxis was implemented through a light-dependence of the swimming speed given by model A  in \cite{Williams2011} (Eq.~\eqref{dndtI}), which should be appropriate mostly when the light intensity $I\approx I_c$.
The model considered also self-shading within the suspension by modifying the light intensity according to the Beer-Lambert law (Eq.~\eqref{Ix}).
Within this model, the microorganisms can be concentrated at specific depths within the suspension by adjusting key parameters like the strength of the flow (defined through the rotational P\'eclet number $Pe = 2U/\left(HD_R\right)$), the cells' absorption coefficient ($\alpha$ in  Eq.~\eqref{Ix}), and the reduced intensity of the light source $I_s/I_c$ (Fig.~\ref{fig:clarke}).       
Unfortunately these studies did not consider the role of bioconvective instabilities due to gravitationally unstable layers induced by photo-focussing in shear flows. We believe that this will provide fertile ground for future investigations.

\section{Conclusion}
\label{sec:conclusion}
Vigorous bioconvective flows  can easily emerge within a suspension of swimming microorganisms, and play a fundamental role in nutrient transport, mixing, and cell distribution within the suspension. This in turn can impact on microbial growth and reproduction, and presents a technological tool to improve the efficiency of bioreactors \cite{Bees2010}. 
We believe that a promising avenue for the rapid and accurate control of bioconvection is to leverage the natural response to light of microbial species.
However, the possibility to engineer photo-bioconvection relies on the ability to predict the phototactic behaviour of individual microorganisms. Unfortunately, our understanding of this process is still largely incomplete. Light-induced steering in microalgae has been investigated at both the physiological and biophysical level \cite{Foster1980, Drescher2010, Arrieta2017,Leptos2018,Demaleprade2019}, and much is known on the mechanism of light detection \cite{Kateriya2004a}, stimulus relay to the flagella \cite{Govorunova1997} and the biomechanics that translates the ensuing changes in flagellar beating into directional changes of the cell \cite{Josef2006,Leptos2018}. 
Upon illumination, however, cells often switch dynamically between positive and negative taxis, even for a fixed light source. 
This seemingly unpredictable behaviour is possibly linked to the dual role of light as both environmental stimulus and energy source, and it highlights the need to investigate in depth the link between cell motility and cell metabolism within a holistic approach to phototaxis.  
We believe this to be an area prime for substantial developments in the future.

Standard microbial photo-bioconvection is currently restricted to motile eukaryotic microalgae, and although several species are technologically important (e.g. {\it Dunaliella}, {\it Chlamydomonas} and {\it Chlorella} spp. \cite{Bees2014}), this still represents a constraint to its general applicability in bioreactors. 
One possible solution is to engineer a light-response within a given species of interest by genetic modification. For example, light-regulation of swimming speed, or photokinesis, has recently been introduced in {\it E. coli}, and used to arrange a bacterial population in complex dynamical patterns of cell concentration in two dimensions \cite{Frangipane2018, Arlt2019}. In turn, this should be able to generate bioconvection within bulk cultures illuminated from above, providing a first example of genetically engineered photo-bioconvection.
Yet another possibility could be offered in the future by light-driven active colloids \cite{Lozano2016,Gomez-Solano2017,Maggi2018}. Currently, they can be made to move microscopic cargo \cite{Ibele2009,Hong2010} and perform phototaxis \cite{Lozano2016} in two dimensions. If their density is sufficiently reduced, they could swim in bulk towards a light source and drive bioconvective instabilities even in suspensions of non-motile microorganisms.

Overall, the possibilities offered by the combination of phototaxis and bioconvection to control the macroscopic behaviour of active matter suspensions are only starting to be explored. We look forward to the new and exciting developments that will undoubtedly emerge in the future.

\acknowledgements
AJ, MP and IT gratefully acknowledge support from The Leverhulme Trust through grant RPG-2018-345. JA and IT acknowledge the support from the Spanish Ministry of Economy and Competitiveness (AEI, FEDER EU) grant nos. FIS2016-77692-C2-1-P and CTM-2017-83774-D. JA thanks the Govern de les Illes Balears for financial support through the Vicen\c{c} Mut subprogram.




\bibliography{photo_bioconvection_bibliography}

\end{document}